\documentclass[twocolumn,prl,superscriptaddress]{revtex4-1} 
\usepackage{graphicx} 
\usepackage{epsfig} 
\usepackage{color} 
\usepackage{braket}
\usepackage{amsmath,amsfonts,amsthm,bm}
\usepackage[breaklinks,colorlinks,linkcolor=blue,citecolor=blue,urlcolor=blue]{hyperref}

\begin{document}

\title{Localization on a synthetic Hall cylinder}
\author{Ren Zhang}
\affiliation{School of Physics, Xi'an Jiaotong University, Xi'an, Shaanxi 710049}
\affiliation{Department of Physics and Astronomy, Purdue University, West Lafayette, IN, 47907}
\author{Yangqian Yan}
\affiliation{Department of Physics and Astronomy, Purdue University, West Lafayette, IN, 47907}
\author{Qi Zhou}
\affiliation{Department of Physics and Astronomy, Purdue University, West Lafayette, IN, 47907}

\date{\today}

\begin{abstract}
By engineering laser-atom interactions, both Hall ribbons and Hall cylinders as fundamental theoretical tools in condensed matter physics have recently been synthesized in laboratories. Here, we show that turning a synthetic Hall ribbon into a synthetic Hall cylinder could naturally lead to localization. Unlike a Hall ribbon, a Hall cylinder hosts an intrinsic lattice, which arises due to the periodic boundary condition in the azimuthal direction, in addition to the external periodic potential imposed by extra lasers. When these two lattices are incommensurate, localization may occur on a synthetic Hall cylinder. Near the localization-delocalization transitions, the dependence of physical observables on the axial magnetic flux allows us to tackle a fundamental question of determining the accuracy of rational approximation of irrational numbers. In the irrational limit, physical observables are no longer affected by fluctuations of the axial flux.
\end{abstract}

\maketitle

Lasers could change the momentum of an atom and meanwhile flip its spin, laying the foundation of synthetic spin-orbit coupling for charge-neutral atoms~\cite{MagField,SOC1,SOC2,SOC3,Livi2016,Kolkowitz,SOCyb,SOCrev,Goldman_2014,Zhai_2015,2dsoc,2dsoc2}. 
Viewing spins as a synthetic dimension~\cite{Celi_2014,Anisimovas}, a one-dimensional spin-orbit coupled system is equivalent to a two-dimensional quantum Hall ribbon~\cite{LENS-ribbons,Stuhl1514}. Synthetic dimensions also allow physicists to access high-dimensional physics~\cite{Ozawa}, such as a four-dimensional charge pump~\cite{Zilberberg,Lohse}. Another profound advantage is the controllable boundary condition~\cite{Boada_2015}. Several groups have recently realized synthetic Hall cylinders by implementing periodic boundary conditions in the synthetic dimensions~\cite{Seoul,Purdue,NIST}. In such a synthetic Hall cylinder, in addition to a uniform axial synthetic magnetic flux, $\phi$, an extra magnetic flux, $\Phi$, can be created at one end surface. Consequently, the total magnetic flux through the cylindrical surface becomes finite, 
a challenging task for a cylinder in real space such as a nanotube. 

Two different scenarios have been used in synthetic Hall ribbons and synthetic Hall cylinders. In experiments done at Florence, Seoul and NIST~\cite{LENS-ribbons,Stuhl1514,Seoul}, 
extra lasers have imposed external optical lattices in the real dimensions. Hamiltonians are constructed based on these external optical lattices, However, in the experiment done at Purdue~\cite{Purdue} and another one at NIST~\cite{NIST}, no external lattice is applied. The density modulation developed on the Hall cylinder purely comes from the periodic boundary condition in the synthetic dimension. For convenience, this is referred to as the intrinsic lattice of the Hall cylinder. Fundamental questions arise. Whether the interplay between the intrinsic and the external lattice leads to any qualitatively new physics? Whether the Hall ribbon and the Hall cylinder could host distinct quantum phenomena? 

The main results of this Letter are summarized as follows. (I) Localization could occur on a Hall cylinder when the external lattice is incommensurate with the intrinsic one. 
(II) It is well-known that any irrational number can be approximated by a rational number to arbitrary precision. This fact manifests itself in our system such that the localization is observable in a finite Hall cylinder with commensurate lattices. The size of the system is a natural cutoff and physical observables of commensurate lattices could well reproduce those of incommensurate lattices. (III) The dependence of physical observables, for instance, the inverse participation ratio, on $\phi$ quantitatively traces how good the rational approximation is. For any commensurate lattices, such dependence is maximized near the delocalization to localization transitions. With increasing the precision of the approximation, the dependence becomes weaker and eventually vanishes in the irrational limit. (IV) Uploading a Bose-Einstein condensate to the Hall cylinder, a weak repulsive interaction favors delocalization but does not change the qualitative physics. (V) None of the above phenomena exists in a Hall ribbon, reflecting the significance of the boundary conditions in synthetic spaces.

\begin{figure}
\centering
\includegraphics[width=0.45\textwidth]{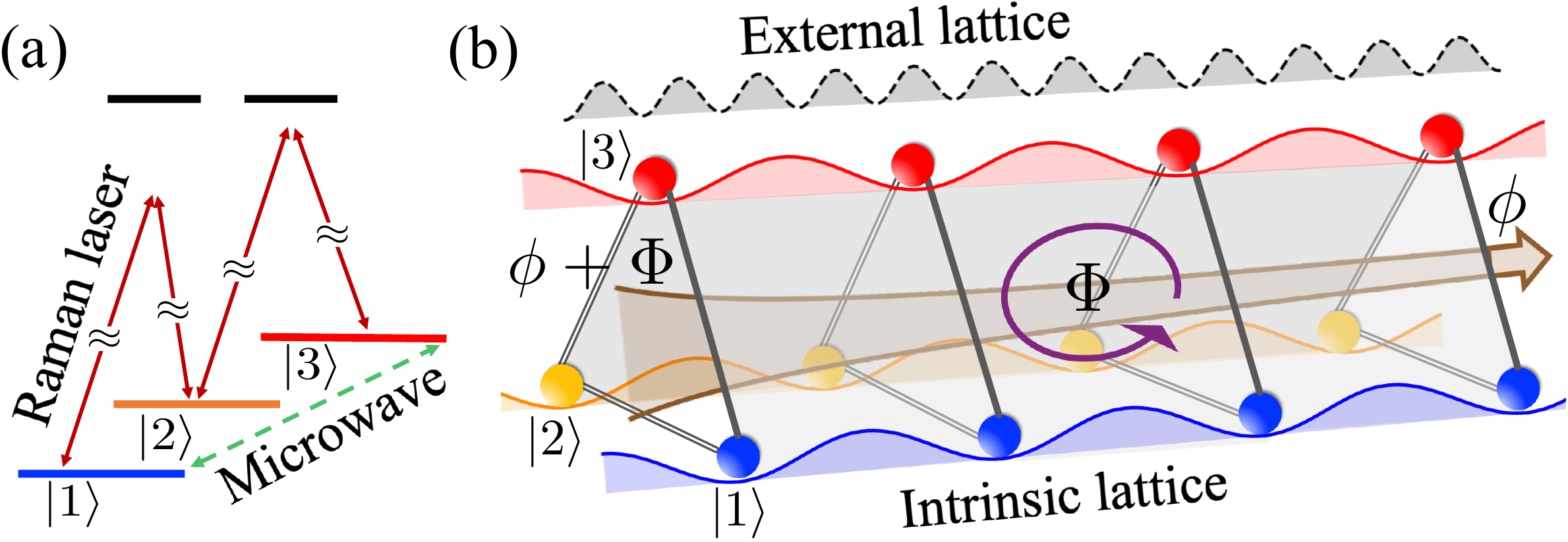}
\caption{A schematic of a synthetic Hall cylinder formed by three internal states. (a): The same Raman lasers couple $|\psi_{1}(x)\rangle$ and $|\psi_{2}(x)\rangle$, $|\psi_{2}(x)\rangle$ and $|\psi_{3}(x)\rangle$. The coupling between $|\psi_{3}(x)\rangle$ and $|\psi_{1}(x)\rangle$ is provided by a two-photon microwave transition. (b): In the real dimension, an external optical lattice is imposed, in addition to the intrinsic lattice on a Hall cylinder. The synthetic flux at the two ends of the Hall cylinder are $\phi$ and $\phi+\Phi$, respectively, where $\Phi$ is equal to the flux penetrating the cylindrical surface.\label{fig1}}
\end{figure}

{\it Hamiltonian and eigenstates ---} 
The synthetic dimension can be encoded in internal states, either hyperfine spins of alkali-metal atoms~\cite{Stuhl1514,Seoul,Purdue,NIST} or nuclear spins of alkaline-earth metal atoms~\cite{LENS-ribbons}. We consider $N$ sites ( $N$ internal states) in the synthetic dimension. As shown in Fig.~\ref{fig1}(a), the inter-site hopping is accomplished by Raman lasers or microwave fields fliping hyperfine or nuclear spins. In the real dimension, extra counter-propagating lasers impose an external one-dimensional optical lattice, the lattice spacing of which is $\pi/k_L$. $k_{L}$ is the wavenumber of the extra lasers. The Hamiltonian is written as
\begin{align}
\label{h0}
\hat{H}_{0}&=\sum_{j=1}^{N}|\psi_{j}(x)\rangle\left[-\frac{\hbar^{2}}{2M}\partial_{x}^{2}+V\sin^{2}(k_{L}x)+\epsilon_{j}\right]\langle\psi_{j}(x)|\nonumber\\
&+\sum_{j=1}^{N}\frac{\Omega_{j}}{2}e^{2ik_{j}x+i\phi_{j}}|\psi_{j+1}(x)\rangle\langle\psi_{j}(x)|+{\rm H.c.},
\end{align}
where $\psi_{j}(x)$ is the spatial wave function of the $j$th internal state, $V$ the depth of the optical lattice, $\epsilon_{i}$ the detuning in the Raman or microwave transition. $\Omega_{j}$ denotes the coupling strength between internal state $j$ and $j+1$ induced by the Raman lasers or a microwave with a wave number $k_{j}$ and phase $\phi_{j}$, and $\psi_{N+1}(x)=\psi_{1}(x)$. $k_j=0$ if it is induced by the microwave. 

Along a closed-loop, $(x_{0},j)\to(x_{0},j+1)\to(x_{0}-\Delta
x,j+1)\to(x_{0}-\Delta x,j)\to(x_{0},j)$, an atom accumulates an extra phase
$2k_{j}\Delta x$. Thus, this system is analog of the 2D electron gas in the
magnetic field. The total magnetic flux penetrating the cylindrical surface,
$\Phi=\sum_j 2k_j L$, where $L$ is the length of the cylinder, originates from
the nonuniform axial flux. As shown in Fig.~\ref{fig1}(b), the axial flux includes a uniform part, $\phi$, and a gradient that leads to different flux through the two ends, $\phi$ and $\phi+\Phi$, respectively. The periodic boundary condition (PBC) is realized by turning on $\Omega_{N}$, the coupling between the first internal state and the $N$th one. The open boundary condition (OBC) is accessed if $\Omega_{N}=0$. For PBC or OBC, we refer to this system as a synthetic Hall cylinder or synthetic Hall ribbon, respectively. On the Hall cylinder, $H_0$ gives rise to an intrinsic periodic lattice even when $V=0$~\cite{Qizhou}.

Hereafter we consider a minimum Hall cylinder ($N=3$). Furthermore, we consider that the coupling between $|\psi_{1}(x)\rangle$ and $|\psi_{2}(x)\rangle$ is provided by the same Raman lasers that couple $|\psi_{2}(x)\rangle$ and $|\psi_{3}(x)\rangle$. $|\psi_{3}(x)\rangle$ and $|\psi_{1}(x)\rangle$ are coupled by a two-photon microwave transition, the strength of which is taken equal to the Raman couplings. As a result, in such a configuration $k_{1}=k_{2}=k_{0}, k_{3}=0$, and $\Omega_{j=1,2,3}=\Omega$. In practice, this configuration can be realized in current experiments~\cite{Seoul,Purdue}. Each single $\phi_j$ can be gauged away but the sum of all phases, $\phi=\sum_j \phi_j$, which represents the constant part of the axial magnetic flux, as shown in Fig.~\ref{fig1}(b), is an important quantity to control system. For convenience, we set $\phi_1=\phi_2=0$, $\phi_3=\phi$.

The periodicity of the density modulation in the intrinsic lattice is $d_{I}=\pi/(2k_{0})$, and the lattice spacing of the external optical lattice is $d_{E}=\pi/k_{L}$. We use $d_{T}$ to denote the lattice spacing of the composite lattice that consists of the intrinsic and external lattices. $d_{T}$ is the least common multiple (lcm) of $d_{I}$ and $d_{E}$, 
\begin{align}
d_{T}={\rm lcm}(d_{I},d_{E}).
\end{align}
We consider a sequence of commensurate lattices that satisfy
\begin{equation}
\left(\frac{d_I}{d_E}\right)_l=\left(\frac{k_L}{2k_0}\right)_l=\frac{f_{l+1}}{2f_l},
\end{equation}
where $f_l=\{1, 1, 2, 3, 5, 8,\cdots\}$ is the Fibonacci sequence with $l=1,2,3,\cdots$ and $f_{l+2}=f_{l+1}+f_{l}$ is its $(l+2)$th element. It is known that $\lim_{l\to\infty}f_{l+1}/f_{l}=(\sqrt{5}+1)/2$, i.e., in the large $l$ limit, a rational number, $f_{l+1}/f_l$, well approximates the golden ratio. 

\begin{figure}
\centering
\includegraphics[width=0.45\textwidth]{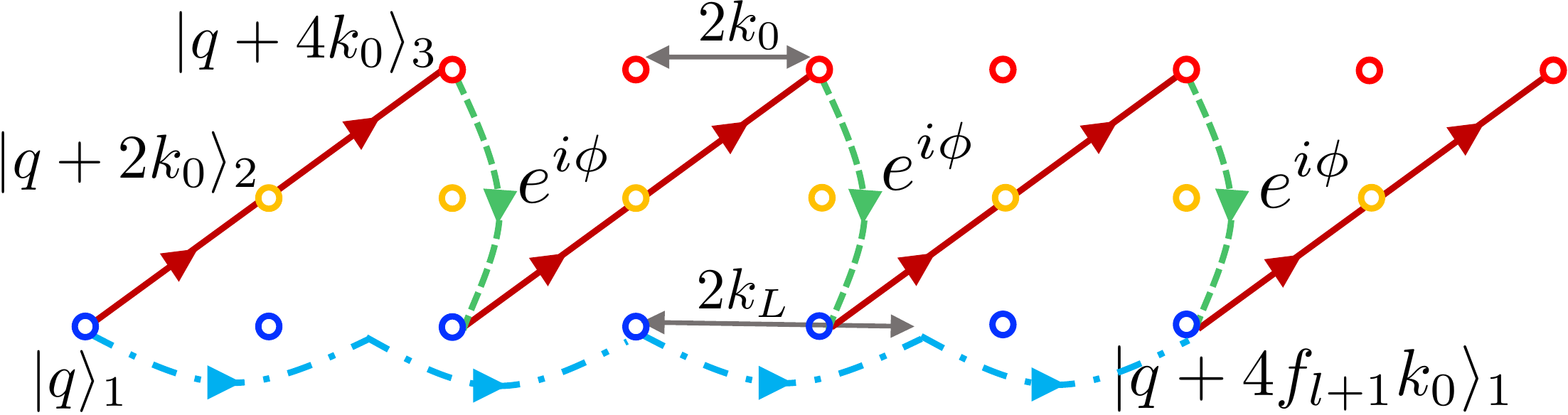}
\caption{Illustration of coupling pathways in the momentum space. One pathway is provided by the Raman (red solid arrow) and microwave (green dashed) transitions, which transfers momentum, $4k_{0}$, and accumulate a phase, $\phi$. Another pathway is provided by the external lattice (blue dash-dotted) which transfers momentum, $2k_{L}$. After $f_{l+1}$ times Raman and microwave transition or $2f_{l}$ times external lattice coupling, the two pathways meet. Here we take $l=3$ as an example.\label{fig2}}
\end{figure}

Figure~\ref{fig2} shows how $\hat{H}_0$ couples a sequence of plane waves, $|q_{j}+4mk_0+2nk_L\rangle_j$, where $m$ and $n$ are integers and $q_{j}=q+2(j-1)k_{0}$. A plane wave of the $j$th spin state, $|q_{j}\rangle_j$, has a onsite energy of $\epsilon_j+\hbar^2q_{j}^2/(2M)$. The inter-site couplings of this momentum space lattice are provided by Raman couplings, microwave transitions and the external lattice, as specified in Fig.~\ref{fig2}. An eigenstate is written as 
\begin{equation}
|\Psi(q)\rangle=\sum_{j,m,n}c_{j,m,n}|q_{j}+4mk_0+2nk_L\rangle_j.
\end{equation}
When $2k_0/k_L$ or equivalently, $d_E/d_I$, is commensurate, two plane waves are coupled by multiple pathways. For instance, Fig.~\ref{fig2} shows two pathways coupling $|q\rangle_1$ and $|q+4f_{l+1}k_0\rangle_1$, one given by the intrinsic lattice and the other by the external lattice. As shown later, the interference between such pathways become crucial when discussing the delocalization to localization transition.

{\it Delocalization to localization transition ---} 
Localization is often studied using the Aubry-Andr\'e (AA) model, which considers particles subject to a random potential~\cite{AA}. In ultracold atoms, a quasi-periodic lattice was realized by imposing two incommensurate optical lattices to mimic the randomness~\cite{roberto,Damski,Biddle}, and the delocalization to localization transition has been observed in bosons~\cite{exp1,exp2}. Here, we show that the delocalization to localization transition naturally emerges on a Hall cylinder. 

We consider a prototypical irrational number, $k_{L}/ k_{0}=(\sqrt{5}+1)/2$, which can be approximated by the ratio of two rational numbers, i.e., the aforementioned $f_{l+1}/f_{l}$, Fibonacci sequence. Such a fact manifests itself in our system in the large $l$ limit, where results of incommensurate lattices can be well approximated by commensurate lattices in a finite system. Strictly speaking, localization exists only for incommensurate lattices, i.e., when $l\rightarrow \infty$. For any finite $l$, $(k_{L}/k_{0})_l=f_{l+1}/f_l$, the external and intrinsic lattices are commensurate. However, when $l$ is large enough, a length scale separation exits, $\sigma\ll L\ll d_T$, where $\sigma$ is the width of the wavefunction in a unit cell of the composite lattice and $L$ is the size of the system. Experiments conducted on such a finite system will well reproduce physics observables in an incommensurate lattice, such as the density distribution.

To quantify the localization, we consider the dimensionless inverse
participation ratio (IPR)~\cite{IPR},
\begin{align}
{\rm IPR}=\frac{\int_{0}^{d_{T}} dx\left[\sum_{j=1}^{3}|\psi_{j}(x)|^{2}\right]^{2}}{k_{0}\left[\int_{0}^{d_{T}} dx\sum_{j=1}^{3}|\psi_{j}(x)|^{2}\right]^{2}}.
\end{align}
For convenience, we have used the total density of all three spin states to compute IPR. In the ideal delocalized states, i.e., $|\psi_{j}(x)|^{2}$ is uniform, IPR $\propto 1/d_{T}$, which approaches zero in the large $l$ limit. In the ideal localized states, i.e., $|\psi_{j}(x)|^{2}=\delta(x)$, IPR diverges. In a finite system, IPR remains small in the delocalized state. Tuning $V$ and $\Omega$, the state become localized and IPR quickly increases to a large value. 
\begin{figure}
\centering
\includegraphics[width=0.45\textwidth]{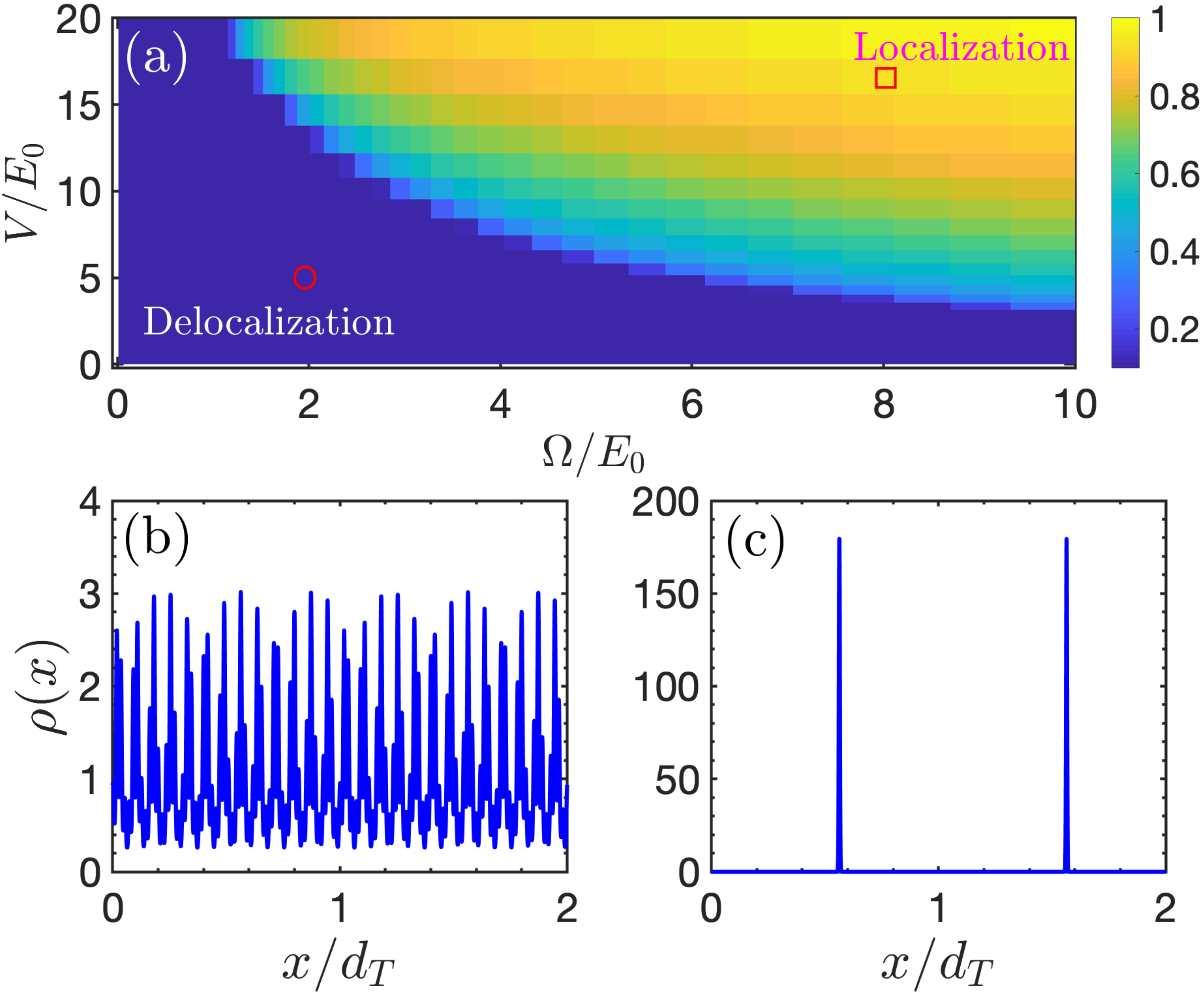}
\caption{(a): Inverse participation ratio (IPR) of the synthetic Hall cylinder. The color bar denotes the density of the dimensionless IPR. (b): A typical ground state density distribution, $\rho(x)=\sum_{j=1}^{3}|\psi_{j}(x)|^{2}$, in the delocalized regime (red circle), where $\Omega=2E_{0}$, $V=5E_{0}$. (c): A typical ground state density distribution in the localized regime 
(red rectangle), where $\Omega=8E_{0}$, $V=18E_{0}$. The density distribution in two unit cells is shown. In our calculation, $\epsilon_{1}=-0.1E_{0}$, $\epsilon_{2}=0$, $\epsilon_{3}=0.1E_{0}$, $\phi=\pi/3$ and $l=9$.\label{fig3}}
\end{figure}

A density plot of IPR as a function of $\Omega$ and $V$ is illustrated in Fig.~\ref{fig3}(a). It is obvious that when either $\Omega$ or $V$ is fixed and the other increases, IPR increases. For instance, when $\Omega=2E_{0}$ and $V=5E_{0}$, as shown by the red circle, IPR $\sim0.1$, and the density distribution is shown in Fig.~\ref{fig3}(b). It is clear that the density distribution extends to the whole unit cell of the composite lattice. In contrast, when $\Omega=8E_{0}$ and $V=18E_{0}$, as shown by the red rectangle, IPR is around the unit, and the density distribution is shown in Fig.~\ref{fig3}(c). The density distribution now is spatially localized in each unit cell, i.e. $\sigma\ll d_T$. More interestingly, there is a clear boundary between the delocalized regime and the localized regime on the density plot of IPR. Near the boundary, IPR increases very quickly from a small value to a large one, which implies the transition from the delocalized state to the localized state. In our calculation, we take $l=9$ meaning $k_{L}/k_{0}=55/34$. Thus, the composite lattice is readily a good approximation for an incommensurate lattice in a finite system, the size of which satisfies $\sigma\ll L\ll d_T$. Whereas it is time-consuming to compute even larger $l$, it is reasonable to expect that with increasing $l$ such that $k_{L}/k_{0}\to(\sqrt{5}+1)/2$, the transition becomes even sharper, which can be served as a smoking gun of the delocalization to localization transition.

{\it Dependence on the uniform axial synthetic flux} --- According to Eq.~(\ref{h0}), the wave function accumulates a constant phase factor $\phi$ after a closed-loop motion $|\psi_{1}(x)\rangle\to|\psi_{2}(x)\rangle\to|\psi_{3}(x)\rangle\to|\psi_{1}(x)\rangle$. In practice, the phases of the Raman lasers are in general different from those of the microwaves. $\phi$ is thus finite and tunable. This corresponds to the uniform part of the axial flux and determines the origin of the intrinsic lattice. In the absence of the external lattice, physical observables in an infinite system do not depend on $\phi$ because of the translational invariance, though introducing a time-dependent $\phi(t)$ could lead to a topological charge pump in a finite cylinder. However, when the external lattice is turned on, $\phi$ cannot be ignored as it determines the relative displacement between the intrinsic and external lattices. Alternatively, one could consider the Hamiltonian in the momentum space, as shown in Fig.~\ref{fig2}. When $k_L/k_0$ is a rational number, e.g. $f_{l+1}/f_l$, multiple pathways contribute to the occupancy in a single plane wave. The uniform flux $\phi$, which determines the relative phase between different pathways, becomes crucial, in particular near the delocalization to localization transition. Such fact allows one to quantitatively trace how good the rational approximation of an irrational number is.

\begin{figure}
\centering
\includegraphics[width=0.45\textwidth]{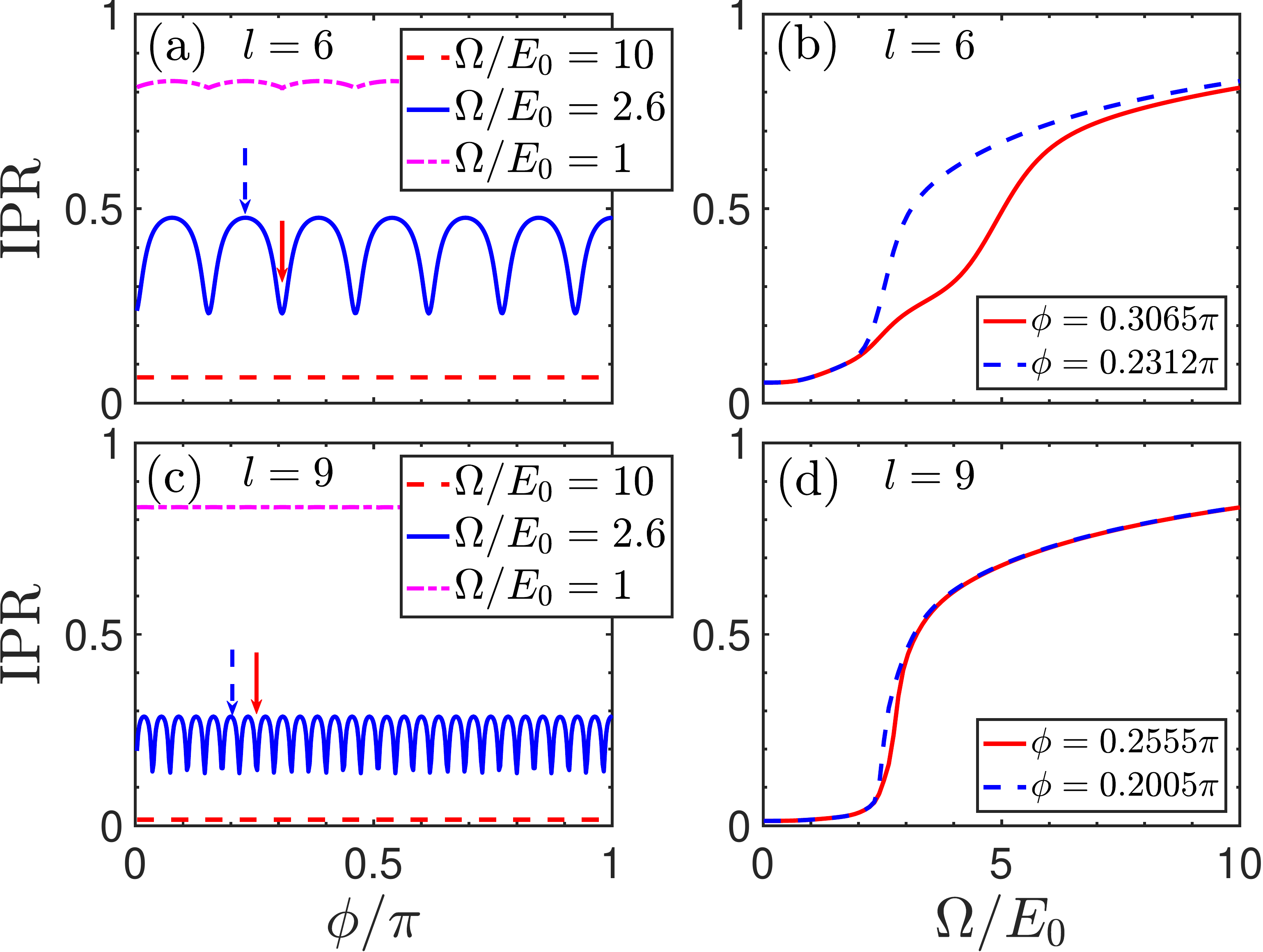}
\caption{(a) and (c): Dependence of IPR on the constant part of the axial synthetic flux, $\phi$, for two rational approximations. $f_{l+1}$ counts the oscillations. (b) and (d): IPR as a function of Raman coupling strength for given external lattice and flux. The values of $\phi$ are indicated by the arrows in (a) and (c). In our calculation, $V=10E_{0}$, $\epsilon_{1}=-0.1E_{0}$, $\epsilon_{2}=0$, $\epsilon_{3}=0.1E_{0}$.\label{fig4}}
\end{figure}

To this end, we calculate the dependence of IPR on $\phi$. Our results indicate that for any commensurate lattice, i.e., a finite $l$, IPR, in general, depends on $\phi$, as expected. However, in both the delocalized state and the localized state, such dependence is weak, as either the intrinsic or the external lattice is much stronger than the other one. Shifting their relative position by changing $\phi$ does not lead to considerable changes in IPR, as shown in Fig.~\ref{fig4}. In contrast, the dependence
is maximized near the transition point, where the amplitudes of these two lattices become comparable. Changing the relative displacement thus significantly changes the property of the system. The same conclusion can be obtained by considering the ``lattice model" in the momentum space. When the amplitudes of the intrinsic and external lattice are comparable to each other, the interference of two different pathways becomes sensitive to their relative phase.

It is worth pointing out that our results are particularly useful in practice, when $\phi$ may fluctuate in repeated experiments. Our findings indicate that near the delocalization to localization transitions, fluctuations in $\phi$ may lead to a large variance of physical observables. If we examine IPR more carefully near the transition point, its periodic dependence on $\phi$ is shown in Fig.~\ref{fig4}(a) and (c). Interestingly, such periodicity directly unfolds $l$ we used in the Fibonacci sequence. Consider a change in $\phi$, $\phi\rightarrow\phi+\Delta \phi$. $\Delta\phi$ can be gauged way by multiplying $|q_{j}+4mk_0+2nk_L\rangle_j$ an extra phase, $e^{im\Delta\phi}$. Thus, $|q_{j}+4f_{l+1}k_0\rangle_j$, or equivalently, $|q_{j}+4f_{l}k_L\rangle_j$, acquires additional phase, $e^{i f_{l+1}\Delta\phi}$, after the transformation. Meanwhile, $|q_{j}+4f_{l+1}k_0\rangle_j$ are coupled to $|q\rangle_j$ by the external lattice potential, resulting in the same additional phase $e^{i f_{l+1}\Delta\phi}$ acquired by $|q_{j}\rangle_j$. To ensure the single-valueness of the wavefunction, $e^{i f_{l+1}\Delta\phi}=1$ has to be satisfied. The periodicity of a physical observable, such as IPR, as a function of $\phi$, is given by $2\pi/f_{l+1}$. We thus conclude that measuring the dependence of IPR or other physical observables on $\phi$ directly allows us to quantitatively trace the precision of using rational numbers to approximate an irrational number.

Another notable result arises when
increasing the precision of the rational approximation of $(\sqrt{5}+1)/2$. With increasing $l$, the dependence of IPR on $\phi$ weakens and eventually vanishes in the incommensurate lattice. For $l=9$, the dependence is less obvious than that of $l=6$, as shown in Fig.~\ref{fig4}(b) and (d). Increasing $l$ means enlarging the length of the pathways that interfere with each other in Fig.~\ref{fig2}. Since $|q_{j}+4f_{l+1}k_0\rangle_j$ has a larger onsite energy than $|q_{j}\rangle_j$, larger $f_{l+1}$ correspond to a weaker coupling between these two states. As a result, the relative phase of these two pathways becomes less important. In particular, in the limit $l\rightarrow \infty$, these two pathways will never meet, as the incommensurate intrinsic and external lattices could not couple $|q_{j}\rangle_j$ to the same plane wave at all. We thus conclude that
the dependence vanishes in the incommensurate lattice. This phenomenon has been
observed in a recent experiment at NIST~\cite{NISTexp}. Our incommensurate lattice corresponds to the irrational transverse flux that suppresses decoherence caused by fluctuations in the axial flux in the experiment.

\begin{figure}
\centering
\includegraphics[width=0.4\textwidth]{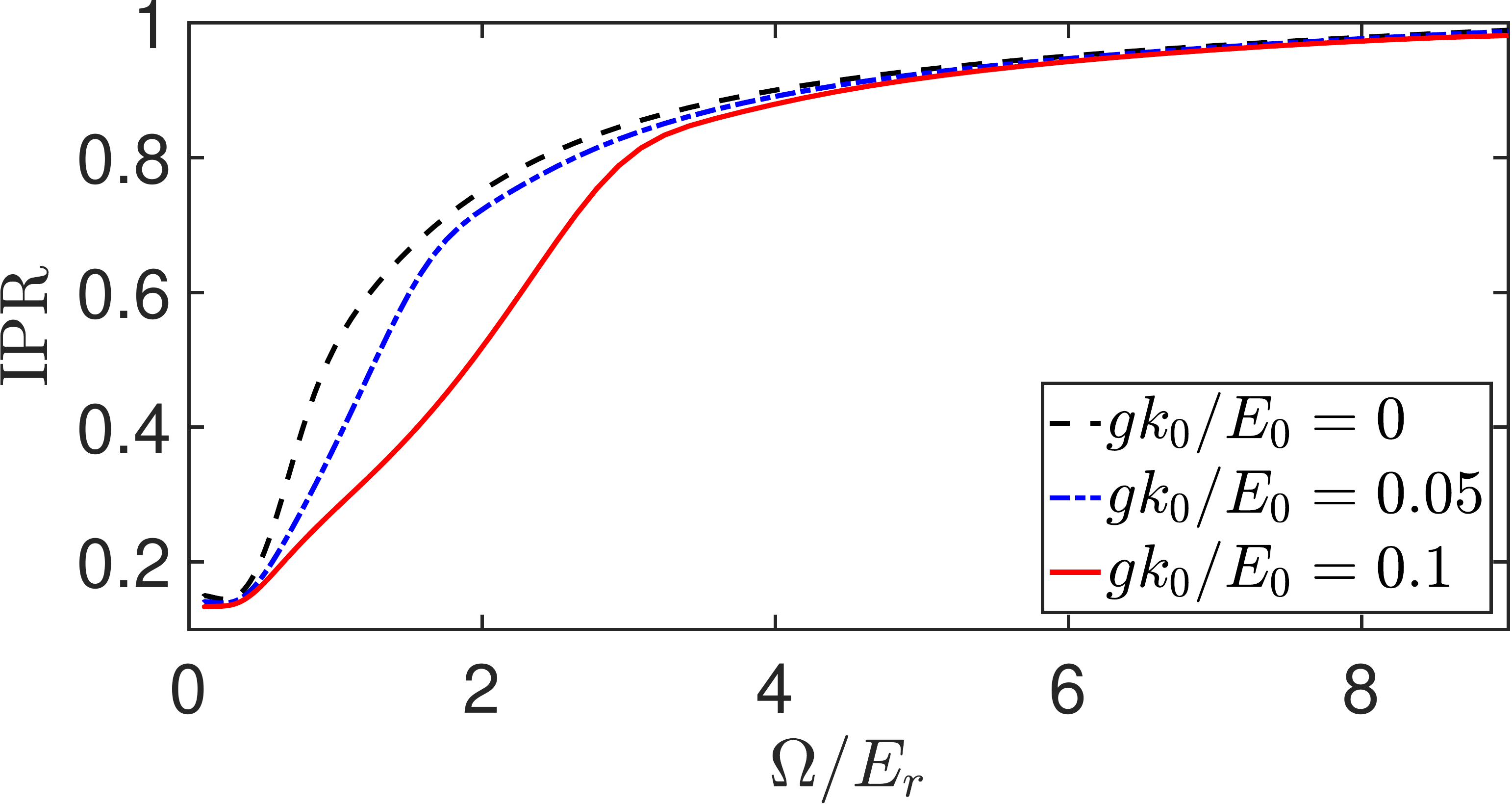}
\caption{Interaction effect on IPR. In our calculation, $V=20E_{0}$, $\hbar\omega=0.05E_{0}$, $\epsilon_{1}=-0.1E_{0}$, $\epsilon_{2}=0$, $\epsilon_{3}=0.1E_{0}$, $l=7$.\label{interaction}}
\end{figure}

{\it Interaction effects ---} In realistic experiments, weak interactions exist in a Bose-Einstein condensate uploaded to a synthetic Hall cylinder. 
For simplicity, we consider equal and repulsive inter-spin and intra-spin interactions, $U=\frac{g}{2}\int dx \rho(x)^2$
where $\rho(x)=\sum_{j=1}^{3}\psi^{\dagger}_{j}(x)\psi_{j}(x)$ denotes the atomic density, $g$ the interaction strength. We also include the harmonic trap in the real dimension. By numerically solving the imaginary-time Gross-Pitaevskii equation,
\begin{align}
-\frac{\partial\vec{\psi}(x)}{\partial\tau}=\left(\hat{H}_{0}+\frac{1}{2}M\omega^{2}x^{2}+g\rho(x)\right)\vec{\psi}(x)
\end{align}
with $\vec{\psi}(x)=(\psi_{1}(x),\psi_{2}(x),\psi_{3}(x))$ and $\omega$ the trapping frequency, we obtain the ground state and hence the IPR. In Fig.~\ref{interaction}, we show the IPR for various interaction strength. Our results shows the a weak repulsive interaction in current experiments slightly changes the IPR but the qualitative features of the non-interacting systems remain.

{\it Quantum Hall ribbon ---} When OBC is applied, the intrinsic lattice
vanishes. Without the external lattice, the density becomes uniform in the real
dimension~\cite{Qizhou}. Turning on a finite $V$, though the density becomes periodically modulated, because of the absence of intrinsic lattice, all previously discussed phenomena regarding the delocalization to localization transition on a Hall cylinder disappear.

In summary, we have studied the delocalization to localization transition on a Hall cylinder due to the interplay between the intrinsic lattice and the external lattice. Our results provide experimentalists with a unique platform to study new quantum phenomena and to address fundamental questions in synthetic spaces. We hope that our work will stimulate the community to explore more synthetic spaces.

\begin{acknowledgments}
Q.Z. acknowledges a useful conversation with Ian Spielman at Sant Feliu that inspired us to expand discussions on the constant part of the axial flux. We are also grateful to Ian Spielman and Qi-Yu Liang to share their manuscript prior to submission. R.Z. is supported by the National Key R$\&$D Program of China (Grant No. 2018YFA0307601), NSFC (Grant No.11804268). Q.Z. and Y. Y. are supported by National Science Foundation (NSF) through Grant No. PHY-1806796, the Air Force Office of Scientific Research under award number FA9550-20-1-0221, and a seed fund from Purdue Quantum Science and Engineering Institute.
\end{acknowledgments}

\end{document}